%% file: manuscript.tex
\documentclass[%
11pt,
reprint,
onecolumn,
tightenlines,
superscriptaddress,
preprintnumbers,
nofootinbib,
amsmath,amssymb,amsthm,
eqsecnum,,
]{revtex4-2}

\input{preamble}

\begin{document}

\preprint{Physical Review E 109:044501, 2024  (\url{https://link.aps.org/doi/10.1103/PhysRevE.109.044501}).}

\title{Statistical Field Theory of Polarizable Polymer Chains with Nonlocal Dipolar Interactions}

\author{Pratik Khandagale\orcidlink{0000-0002-3105-1439}}
    \email{pkhandag@alumni.cmu.edu}
    \affiliation{Department of Mechanical Engineering, Carnegie Mellon University}

\author{Carlos Garcia-Cervera}
    \affiliation{Mathematics Department, University of California, Santa Barbara}
    \affiliation{BCAM, Basque Center for Applied Mathematics, E48009 Bilbao, Basque Country, Spain}

\author{Gal deBotton}
    \affiliation{Department of Mechanical Engineering, Ben Gurion University}
    \affiliation{Department of Biomedical Engineering, Ben Gurion University}
    
\author{Timothy Breitzman}
    \affiliation{Air Force Research Laboratory}
    
\author{Carmel Majidi}
    \affiliation{Department of Mechanical Engineering, Carnegie Mellon University}
    \affiliation{Department of Civil and Environmental Engineering, Carnegie Mellon University}
    \affiliation{Department of Materials Science and Engineering, Carnegie Mellon University}

\author{Kaushik Dayal\orcidlink{0000-0002-0516-3066}}
    \affiliation{Department of Civil and Environmental Engineering, Carnegie Mellon University}
    \affiliation{Department of Mechanical Engineering, Carnegie Mellon University}
    \affiliation{Center for Nonlinear Analysis, Department of Mathematical Sciences, Carnegie Mellon University}
    
\date{\today}

\begin{abstract}
    The electromechanical response of polymeric soft matter to applied electric fields is of fundamental scientific interest as well as relevant to technologies for sensing and actuation.
    Several existing theoretical and numerical approaches for polarizable polymers subject to a combined applied electric field and stretch are based on discrete monomer models.
    In these models, accounting for the interactions between the induced dipoles on monomers is challenging due to the nonlocality of these interactions.
    On the other hand,  the framework of statistical field theory provides a continuous description of polymer chains that potentially enables a tractable way to account for these interactions.
    However, prior formulations using this framework have been restricted to the case of weak anisotropy of the monomer polarizability.
    
    This paper formulates a general approach based in the framework of statistical field theory to account for the nonlocal nature of the dipolar interactions without any restrictions on the anisotropy or nonlinearity of the polarizability of the monomer.
    The approach is based on 3 key elements:
    (1) the statistical field theory framework, in which the discrete monomers are regularized to a continuous dipole distribution;
    (2) a replacement of the nonlocal dipole-dipole interactions by the local electrostatics PDE with the continuous dipole distribution as the forcing;
    (3) the use of a completely general relation between the polarization and the local electric field. 
    Rather than treat the dipole-dipole interactions directly, the continuous description in the field theory enables the computationally-tractable nonlocal-to-local transformation.
    Further, it enables the use of a realistic statistical-mechanical ensemble wherein the average far-field applied electric field is prescribed, rather than prescribing the applied field at every point in the polymer domain. 
    
    The model is applied, using the finite element method (FEM), to study the electromechanical response of a polymer chain in the ensemble with fixed far-field applied electric field and fixed chain stretch.
    The nonlocal dipolar interactions are found to increase, over the case where dipole-dipole interactions are neglected, the magnitudes of the polarization and electric field by orders of magnitude as well as significantly change their spatial distributions.
    Next, the effect of the relative orientation between the applied field and the chain on the local electric field and polarization is studied.
    The model predicts that the elastic response of the polymer chain is linear, consistent with the Gaussian approximation, and is largely unchanged by the orientation of the applied electric field, though the polarization and local electric field distributions are significantly impacted.
    
\end{abstract}

\maketitle

\section{Introduction}

Stimuli-responsive polymeric soft matter is central to actuators and sensors in applications such as soft robotics
\cite{brochu2012dielectric,kim2007electroactive,rosset2013self,leroy2020multimode,ji2019autonomous,cacucciolo2019stretchable,ware2016localized,babaei2021torque,huang2012giant, erol2019microstructure,logrande2023dimensionally}, stretchable electronics \cite{rogers2010materials,kim2009large,khang2006stretchable,xu2013stretchable}, energy harvesting \cite{mckay2014dielectric,han2019three,erturk2011piezoelectric,nan2018compliant}, healthcare \cite{kwak2020wireless,zhang2020water,xu2014soft,kang2016bioresorbable,kim2011epidermal}, and functional systems broadly
\cite{jiao2014understanding,muthukumar2023physics,nepal2023polymer, zolfaghari2020network,amjadi2016stretchable, deng2014electrets,chen2021interplay,zhao2021modeling, liu2013giant,galipeau2013finite,zhang2017nonlinear}.
Electro-responsive polarizable polymers such as dielectric elastomers (DEs) are naturally soft, lightweight, compliant, and can undergo large deformation under an applied electric field, making them promising candidate materials.
However, there are also shortcomings with currently-available polymerics materials, e.g., they often need highly applied electric fields to achieve a meaningful level of actuation \cite{bar2004electroactive}. 
A fundamental understanding of the physics of polymers subjected to electric fields is essential to improve existing, and discover new, polarizable polymeric materials, e.g., in the case of statistical mechanics applied to soft matter flexoelectricity \cite{grasinger2021flexoelectricity,kulkarni2023fluctuations,hassan2023entropic,ahmadpoor2013apparent,ahmadpoor2015flexoelectricity,zelisko2017determining}.

\paragraph*{Prior Work.}

\begin{figure}[htb!]
	\centering
    \includegraphics[width=0.8\textwidth]{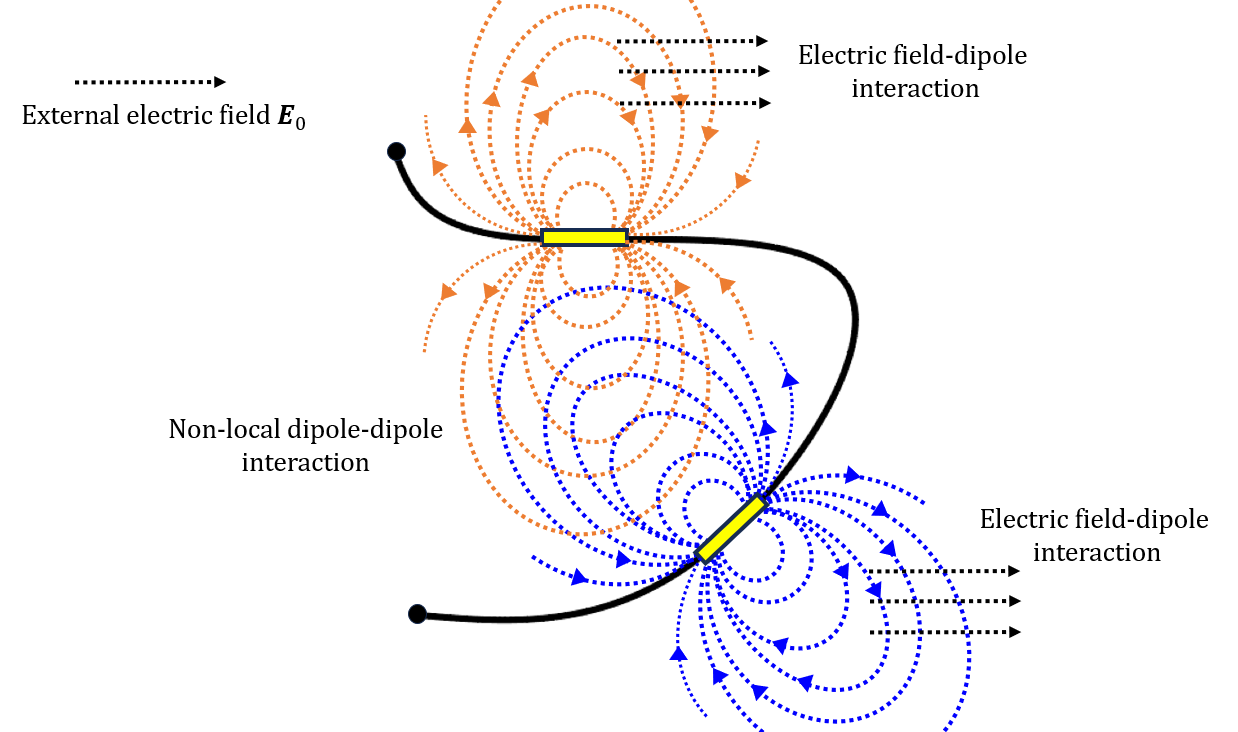}
    \caption{
    	A polarizable polymer chain in an externally-applied far-field electric field $\bfE_0$ induces dipoles in the polymer segments.
    	The dipole-dipole interactions are long range --- they decay as the inverse third power of the distance --- and lead to nonlocal effects that cannot be truncated without very large error \cite{marshall2014atomistic}.
    	These, in turn, affect the configuration of the chain and the mechanical response.
    	We show in this paper that neglecting the dipole-dipole interactions and only accounting for the applied field-dipole interactions, as in prior work, leads to large errors.
    }
\label{fig:intro}
\end{figure}

The physics of polarizable polymeric soft matter is governed by the polymer chain entropy, the interaction between the applied electric field and the induced dipoles, and the nonlocal dipolar interactions between the polymer segments (Fig. \ref{fig:intro}). 
Existing models for electro-responsive polymeric soft matter can be broadly divided into two categories: continuum based-approaches, e.g., \cite{toupin1956elastic, dorfmann2014nonlinear, liu2018emergent, krichen2019liquid, li2015geometrically, darbaniyan2019designing, fox2008dynamic,li2023constitutive,furer2022homogenization},
and statistical mechanics-based approaches, e.g., 
\cite{cohen2016electroelasticity, cohen2016electromechanical, grasinger2020statistical, grasinger2021architected, grasinger2021nonlinear, grasinger2021flexoelectricity, grasinger2022statistical,itskov2019electroelasticity}.

Continuum approaches typically formulate the free energy density by coupling established rubber elasticity models to continuum electrostatics.
These approaches are useful in enabling the study of electro-responsive polymers in complex and realistic geometries and boundary conditions.  
However, these approaches cannot provide predictive insights that are based in the response of the individual monomers. 

The statistical mechanics based-approaches for polymers, on the other hand, are capable of accounting for the molecular details of the polymer chain. 
Statistical mechanics has been employed for several decades to study the mechanical response of polymers and their networks, e.g., \cite{treloar1975physics, kuhn1946statistical, weiner2012statistical, marckmann2006comparison,khandagale2023statistical}.
In the context of electro-responsive polymers, the first works that applied statistical mechanics appear to be \cite{cohen2016electroelasticity} and \cite{cohen2016electromechanical}. 
Broadly, their work derives an approximate expression for the most probable density of monomer orientations that is exact when the polymer chain is not stretched. 
Building on this, a statistical mechanics-based discrete monomer model for an electro-responsive polarizable polymer chain was presented in \cite{grasinger2020statistical,grasinger2022statistical}. 
Using the maximum term approximation assumption, they evaluated the most probable density of monomer orientation and the free energy of polymer chain applicable at large stretches. 
Although these theoretical approaches for electro-responsive polymers provide valuable insights, they all ignore the nonlocal dipolar interactions; they only model the interaction between the applied electric field and the induced dipoles.
A key reason for this limitation is that these approaches all use a discrete description of the polymer chain, and it is computationally very expensive to account for all pairs of interactions.

In contrast, a statistical field theoretic formulation \cite{fredrickson2007computational,fredrickson2002field,fredrickson2006equilibrium} for polarizable polymer chains in an external electric field was presented in \cite{martin2016statistical} that accounts for the dipolar interactions.
A key feature of the field theoretic approach is that the polymer is described as a continuous --- rather than discrete --- object, enabling the replacement of dipole-dipole interaction sums by integrals \cite{shen2017electrostatic,jiang2018density,zhuang2018statistical}.
However, the formulation in \cite{martin2016statistical} is limited to the setting of weak anisotropy of the polymer polarizability.
Also, in the field theoretic formulation, it is challenging to account for realistic electrical boundary conditions, i.e., typically the statistical mechanics ensemble assumes that the applied field is given at every point, rather than only the far-field or average value which is more realistic.

\paragraph*{Contributions of This Work.}

To overcome the limitations of the existing models for polarizable polymeric soft matter, we have developed a statistical field theoretic framework for polarizable polymer chains that enables us to account for the nonlocal dipole-dipole interactions among the polymer segments. 
The approach is based on three elements.

First, the statistical field theory framework, in which the polymer chain is regularized to a continuous description.
We model the polarizable flexible polymer chain using a worm-like chain (WLC) model with field-induced dipoles along the length of the chain.
The continuous description enables us to avoid treating individual interactions between discrete dipoles, but instead as a tractable continuous polarization distribution.

Second, the continuous polarization description enables us to replace the nonlocal dipole-dipole interactions --- an integral operation --- by the local electrostatics partial differential equation (PDE) which uses instead the interaction with the local electric field set up the dipoles.
To compute the effective bound charge, we introduce a polarization operator for a polymer chain, and derive an expression for the thermodynamically-averaged induced polarization. 

Third, we use a completely general dielectric response function that relates the induced polarization to the local electric field. 
While we use a linear anisotropic relation between the polarization and the electric field for the numerical calculations, the method is directly applicable to general nonlinear response functions. 
Further, the dielectric response is necessarily nonlinear in the orientation to satisfy frame invariance.
These elements provide a self-consistent field theoretic formulation to obtain the properties of the polymer chain --- such as segment density, polarization distribution, and local electric field distribution around the polymer chain --- under an externally applied electric field.

A significant aspect of our formulation is that we apply the external electric field only on the boundary of the spatial domain that includes the polymer chain as well as free space.
This corresponds to a thermodynamic ensemble with specified far-field or average applied electric field \cite{grasinger2020statistical}, which is realistic in terms of experimental configurations.
The local electric field is obtained self-consistently through solving for the electrostatic equation that accounts for the polarization distribution.
By using the finite element method (FEM) with an unstructured discretization, we are able to efficiently solve by refining the mesh around the polymer chain where variations are large and keeping it coarse in the free space away from the chain.

The FEM implementation is applied to study the electromechanical response of a polymer chain in the ensemble with fixed far-field applied electric field and fixed chain stretch.
We find that the nonlocal dipolar interactions are found to increase, over the case where dipole-dipole interactions are neglected, the magnitudes of the polarization and electric field by orders of magnitude as well as change significantly their spatial distributions.
Next, we study the effect of the relative orientation between the applied field and the chain on the local electric field and polarization.
When the applied electric field is aligned with the chain end-to-end vector, the larger values of the polarization and electric field are primarily concentrated near the constrained chain ends. 
In contrast, when the applied field is orthogonal to the chain end-to-end vector, the larger values of the polarization and electric field are distributed along the chain. 
However, we observe that despite these differences in the polarization and local field, the elastic response of the chain is linear and largely unchanged by the different orientations of the applied electric field.

\paragraph*{Organization.}

Section \ref{sec:formulation} presents our approach; Section \ref{sec:numerics} presents the numerical method; and Section \ref{sec:results} presents results from our calculations.

\section{Formulation}
\label{sec:formulation}

This section presents the formulation of the framework for a polarizable polymer chain under a far-field applied electric field.
First, we summarize the standard self-consistent statistical field theoretic description of the polymer chain, following \cite{fredrickson2006equilibrium}.
Next, we introduce the polarization operator and its thermodynamic average for the polarizable polymer chain, and describe the coupling between electrostatics and polymer chain description.

\subsection{Self-Consistent Statistical Field Theory Description of a Polymer Chain}

\begin{figure}[ht!]
   	\centering
	\includegraphics[width=0.7\textwidth]{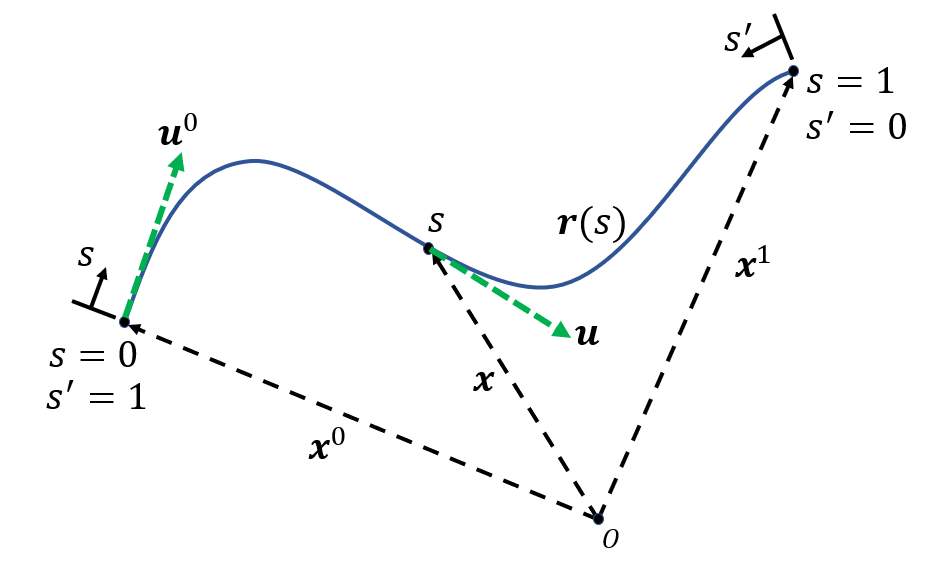}
   	\caption{Kinematic quantities used to describe the polymer chain.}
\label{fig:single_worm_like_chain}
\end{figure}

We use a worm-like chain (WLC) model in this paper with a small but nonzero value for the persistence length.
The chain has $N$ coarse-grained polarizable polymer segments, each with length $a$; $L_c=aN$ is the total contour length of the chain. 
The persistence length is denoted by $\lambda$, i.e., the distance along the polymer chain contour over which the orientational correlations decay. 
The ratio $\lambda/L_c$ determines the flexibility of the chain: $\lambda/L_c \ll 1$ gives a very flexible polymer chain, whereas $\lambda/L_c \gg 1$ gives a rigid rod-like polymer chain.  
To model a flexible polymer chain in this work, we use $\lambda/L_c=10^{-3}$. 

A coarse-grained configuration of the polymer chain is represented as a continuous 3-d space curve $\bfr(s)$ in Figure \ref{fig:single_worm_like_chain}, where $s$ is the chain contour coordinate that is non-dimensionalized with chain contour length $L_c$ such that $0 \leq s \leq 1$. 
The position and orientation of the chain segment with contour coordinate $s$ are given by $\bfx=\bfr(s)$ and  $\bfu=\frac{1}{L_c}\frac{\dm\bfr}{\dm s}$, respectively. 
The chain is assumed to be inextensible, hence $\bfu$ is a unit vector.
The position and orientation at the ends of the chain are denoted $\bfx^0$ and $\bfu^0$ at $s=0$ and  $\bfx^{1}$ and $\bfu^{1}$ at $s=1$.

The fundamental statistical mechanics quantity is the partition function, $Q[w]$, which has the expression:
\begin{equation}\label{eq:Q}
    Q[w]= \frac{1}{4 \pi V } \int \dm \bfx \int \dm \bfu \ q(\bfx, \bfu, \bfx^0, \bfu^0, s) q^{*}(\bfx, -\bfu, \bfx^{1}, \bfu^{1}, 1-s)  ,
\end{equation}
where $w(\bfx, \bfu)$ is the potential of the external field, originating in the interaction between the induced dipole on the polymer segment and the local electric field and whose form is given in \eqref{eq:w_afo_E}; 
$q(\bfx, \bfu, \bfx^0, \bfu^0, s) $ and $q^{*}(\bfx, -\bfu, \bfx^{1}, \bfu^{1}, 1-s) $ are the partial partition functions of the chain for the two chain fragments, one from $0$ to $s$ and the other from $1$ to $s$, respectively (Fig. \ref{fig:single_worm_like_chain});
and $V = Na^2$ is the volume of the polymer chain in 2-d (we would use $V=Na^3$ for 3-d).
The domain of integration is over the spatial domain in $\bfx$ and over the unit sphere in $\bfu$.

The partial partition functions $q$ and $q^{*}$ are obtained by solving the PDEs below \cite{fredrickson2006equilibrium}:
\begin{align}
	\label{eq:q_PDE}
	\frac{\partial q}{\partial s} &= -w(\bfx, \bfu) q -L_c \bfu \cdot \nabla_{\bfx} q + \frac{L_c}{2 \lambda} \nabla^2_{\bfu}q
	\\
	\label{eq:q_star_PDE}
	\frac{\partial q^{*}}{\partial s'} &= -w(\bfx, \bfu) q^{*} - L_c \bfu \cdot \nabla_{\bfx} q^{*} + \frac{L_c}{2 \lambda} \nabla^2_{\bfu}q^{*}
\end{align}
where $s' = 1-s$ varies along the chain contour in the opposite sense as $s$ (Fig. \ref{fig:single_worm_like_chain}). 

The corresponding initial conditions are:
\begin{align}
	\label{eq:q_ini_condition}
	q(\bfx, \bfu, \bfx^0, \bfu^0, s)\Big|_{s=0} &= V \ \delta (\bfx- \bfx^0).
	\\
	\label{eq:q_star_ini_condition}
	q^{*}(\bfx, -\bfu, \bfx^{1}, \bfu^{1}, s')\Big|_{s'=0} &= V \ \delta (\bfx- \bfx^{1})
\end{align}

The initial conditions in \eqref{eq:q_ini_condition} and \eqref{eq:q_star_ini_condition} specify the physical constraint that the ends of the chain are fixed at $\bfx^0$ and $\bfx^{1}$. 
We do not constrain the chain orientations at the ends.

The linear PDEs in \eqref{eq:q_PDE} and \eqref{eq:q_star_PDE} are Fokker–Planck equations that govern the propagation of correlations in segment position and orientation for a worm-like polymer chain under an external field $w(\bfx, \bfu)$.
This system of PDEs is derived using a recursive relation based on the Markov property of the polymer chain partition function.
Physically, these PDE imply that the partition function for a chain fragment from $0$ to $s+\Delta s$ can be composed of two contributions: first, the partition function for the chain from $0$ to $s$, and second, the partition function for a small additional chain segment between $s$ and $s+\Delta s$.
The partition function for this additional segment is then written, using a Boltzmann weight, in terms of the energy that consists of a quadratic contribution from chain bending and the energy due to the external field $w$. 
Taking the limit $\Delta s \to 0$ leads to the system of PDE; for the details of the derivation, we refer to Section 2.5 in \cite{fredrickson2006equilibrium}.

The first term on the right sides of \eqref{eq:q_PDE} and \eqref{eq:q_star_PDE} relate to the external field acting on the polymer chain.
The operator $\nabla_{\bfu}^2 $ in the third term on the right side is the rotational diffusion operator that generates diffusive motion on the unit sphere. 

In general, by using the appropriate form for the functional dependence of $w(\bfx,\bfu)$ on $\bfu$, these equations can be used to describe the interaction between worm-like polymer chains with a broad class of external potentials, including electric and magnetic fields \cite{fredrickson2006equilibrium}. 
Examples include polymer chains with induced or permanent electric or magnetic dipoles along the polymer backbone, and anisotropic potential fields that can model liquid crystalline behavior. 
In this work, we use $w(\bfx,\bfu)$  to model the electrostatic energy of the dielectric polymer chain due to induced electric dipoles in the polymer segments in an externally applied electric field.

\subsection{Spatial Dipole Distribution}

The density operator $\hat{\rho}(\bfx, \bfu)$ for the WLC is defined as \cite{fredrickson2006equilibrium}:
\begin{equation}\label{eq:rho_def}
    \hat{\rho}(\bfx, \bfu):= \int\limits_{0}^{1} \dm s \ \delta(\bfx- \bfr(s)) \delta \left(\bfu- \frac{1}{L_c} \bfr'(s) \right) \delta(|\bfu|-1),
\end{equation}
The Dirac measures build in the kinematic definitions of $\bfx$ and $\bfu$ as constraints. 
The thermodynamically-averaged segment density, $\langle \hat{\rho}(\bfx, \bfu) \rangle$, is then obtained as \cite{fredrickson2006equilibrium}:
\begin{equation}\label{eq:rho_avg}
	\langle  \hat{\rho}(\bfx, \bfu) \rangle = \frac{1}{ 4 \pi V Q[w]} \int\limits_{0}^{1} \dm s \ q(\bfx, \bfu, \bfx^0, \bfu^0, s) q^{*}(\bfx, -\bfu, \bfx^{1}, \bfu^{1}, 1-s).
\end{equation}
where $\langle \cdot \rangle$ corresponds to the statistical average performed over all possible conformations of the polymer chain.

Next, we introduce the polarization operator $\hat{\bfp}(\bfx, \bfu)$ as:
\begin{equation}\label{eq:polarization_operator_x_u}
    \hat{\bfp}(\bfx, \bfu) := 4 \pi V   \int\limits_{0}^{1} \dm s \ \bfp_{seg}(\bfx, \bfu)  \ \delta(\bfx- \bfr(s))  \delta \left(\bfu- \frac{1}{L_c} \bfr'(s) \right) \delta(|\bfu|-1) ,
\end{equation}
where $\bfp_{seg}(\bfx, \bfu)$ is the polarization response function, i.e., the induced polarization at the point $(\bfx, \bfu)$ in configuration space.

Defining $\langle \hat{\bfp}(\bfx, \bfu) \rangle $ as the thermodynamically-averaged polarization of the polymer segment at $(\bfx, \bfu)$ in configuration space, we write: 
\begin{equation}\label{eq:avg_polarization_x_u_appendix}
\begin{split}
    \langle \hat{\bfp}(\bfx, \bfu) \rangle 
    &=  
    \left\langle 4 \pi V    \int\limits_{0}^1 \dm s \ \bfp_{seg}(\bfx, \bfu)  \ \delta(\bfx- \bfr(s)) \delta \left(\bfu- \frac{1}{L_c}\bfr'(s)\right) \delta(|\bfu|-1) \right\rangle \\
    &= 
    4 \pi V \bfp_{seg}(\bfx, \bfu) \left\langle   \int\limits_{0}^1 \dm s \   \ \delta(\bfx- \bfr(s)) \delta \left(\bfu- \frac{1}{L_c}\bfr'(s)\right) \delta(|\bfu|-1) \right\rangle\\
    &= 
    4 \pi V \bfp_{seg}(\bfx, \bfu) \left\langle  \hat{\rho}(\bfx, \bfu)  \right\rangle, \quad \text{using \eqref{eq:rho_def} and \eqref{eq:rho_avg}.}
\end{split}
\end{equation} 

We define $\bfp(\bfx)$, the polarization at the spatial location $\bfx$, as the average over $\bfu$ at the location $\bfx$:
\begin{equation}\label{eq:avg_polarization_x}
  \bfp(\bfx) := \frac{1}{4 \pi} \int \dm \bfu \
            \langle \hat{\bfp}(\bfx, \bfu) \rangle  = V \int \dm \bfu \
            \bfp_{seg}(\bfx, \bfu)  \langle  \hat{\rho}(\bfx, \bfu) \rangle 
\end{equation} 
This final quantity $\bfp(\bfx)$ will appear in the electrostatic equation in the bound charge density.

\subsection{Electrostatics}\label{sec:self_consistent_formulation}

We next obtain the electric field through the local electrostatics PDE, which is tractable numerically since the charge distribution described through $\bfp(\bfx)$ does not involve singular dipole distributions.
This also lets us directly apply realistic boundary conditions --- i.e., specified potential on the boundary of the domain, corresponding to a given far-field applied electric field --- without having to compute the Greens function for a given geometry.
As shown in the Appendix of \cite{grasinger2020statistical}, this ensemble is equivalent to prescribing the average field over the domain.
The interior field within the domain is a superposition of the electrostatic interaction between the applied field and the induced dipole response of the polymer segments as well as the nonlocal dipole-dipole interactions among the induced dipoles.
The usual electrostatics PDE accounts automatically for all of these interactions.

We start from the electrostatic equation for the electrostatic potential $\phi(\bfx)$:
\begin{equation}
    - \epsilon_0 \nabla^2 \phi(\bfx)= -\divergence \bfp \text{ on } \Omega, \quad \text{ given } \phi(\bfx)= -\bfE_0\cdot \bfx \text{ on } \partial\Omega
\end{equation} 
where $\bfp$ is obtained from \eqref{eq:avg_polarization_x}; $-\divergence \bfp$ is the bound charge density; $\bfE_0$ is the given average electric field; and $\Omega$ is the region of space with boundary $\partial \Omega$ over which we solve the electrostatic problem.

The electric field $\bfE(\bfx)$ is related to $\phi(\bfx)$ through the classical relation:
\begin{equation}\label{eq:E_elec_potential_relation}
    \bfE(\bfx)= - \nabla \phi(\bfx).
\end{equation}
In turn, the field $w(\bfx, \bfu)$ is related to $\bfE(\bfx)$ as \cite{fredrickson2006equilibrium}:
\begin{equation}\label{eq:w_afo_E}
    w(\bfx, \bfu)
    =  - \frac{4 \pi V}{2 k_B T} \bfp_{seg}(\bfx,\bfu)\cdot\bfE(\bfx)
\end{equation}

\subsection{Monomer Dipole Response}

We assume that the monomers that compose the chain have a dielectric response that is linear in the electric field\footnote
{
    While we use a dielectric response that is linear in the electric field for this paper to perform explicit numerical calculations, it is equally easy to incorporate more general nonlinear responses.
}.
However, the response is necessarily nonlinear in the orientation $\bfu$ to satisfy rotation invariance of the monomer.

We define the polarization response of a segment of the polymer chain as:
\begin{equation}\label{eq:polarization_linear_form}
	\bfp_{seg}(\bfx, \bfu):= \epsilon_0 \bfbeta(\bfu) \bfE(\bfx).
\end{equation}
where $\bfbeta$ is related to the molecular polarizability tensor of the chain segment, and depends on the orientation of the chain segment $\bfu$.
We model the polarizability tensor $\bfbeta$ as transversely isotropic \cite{cohen2016electromechanical,cohen2016electroelasticity} with the expression:
\begin{equation}\label{eq:linear_polarizability}
	\bfbeta(\bfu)= \beta_{\parallel} \bfu \otimes \bfu + \beta_{\perp} \left(\bfI-\bfu \otimes \bfu\right) .
\end{equation}
where $\beta_{\parallel}$ and $\beta_{\perp}$ are the polarizabilities of the segment along the segment orientation and transverse to the segment orientation, respectively.

\subsection{Model Summary}

The external field $w(\bfx, \bfu)$ connects to the PDEs for the partial partition functions $q$ and $q^*$ in \eqref{eq:q_PDE} and \eqref{eq:q_star_PDE}. 
The solutions for $q$ and $q^*$ in turn relate to the partition function $Q[w]$ and average segment density $\langle \hat{\rho}(\bfx, \bfu)  \rangle $ using \eqref{eq:Q} and \eqref{eq:rho_avg}, respectively. 
Finally, to close the loop, the polarization, $\bfp(\bfx)$, is related by \eqref{eq:avg_polarization_x} to $Q[w]$, $\langle \hat{\rho}(\bfx, \bfu)  \rangle $, and $\bfp_{seg}$.

\section{Numerical Method}
\label{sec:numerics}

\subsection{Self-consistent Iteration}\label{sec:self_consistent_iteration}

\begin{algorithm}[H]
	\caption{Self-consistent iterative algorithm to compute the equilibrium properties of a polymer chain}
	\begin{algorithmic}
		\While{$\Delta Q > \epsilon = 10^{-3}$} 
		\State Compute $ \bfp(\bfx) = V \int \dm \bfu \
		\bfp_{seg}(\bfx, \bfu)  \langle  \hat{\rho}(\bfx, \bfu) \rangle $ \Comment{ $\bfp_{seg}(\bfx, \bfu)= \epsilon_0 \bfbeta(\bfu) \bfE(\bfx)$}
		\State Solve for $\phi(\bfx)$: $\nabla^2 \phi(\bfx)= \frac{1}{\epsilon_0} \divergence  \bfp(\bfx) $, $\quad$ given  $\phi(\bfx)= -\bfE_0\cdot \bfx$ on ${\partial \Omega}$
		\State Compute $\bfE(\bfx)=-\nabla \phi(\bfx)$
		\State Compute $ w(\bfx, \bfu)= -  \frac{4 \pi V}{2 k_B T} \Big[ \epsilon_0 \bfbeta(\bfu) \bfE(\bfx) \Big] \cdot \bfE(\bfx)$
		\State Compute $q$ and $q^{*}$, by solving  \eqref{eq:q_PDE} and \eqref{eq:q_star_PDE}, respectively
		\State Compute $Q[w]$ and $\langle \hat{\rho}(\bfx, \bfu) \rangle$, using \eqref{eq:Q} and \eqref{eq:rho_avg}, respectively
		\EndWhile
		\State \textbf{Outputs:} $ \bfE^{eq}(\bfx), Q^{eq}, \langle \hat{\rho}(\bfx, \bfu) \rangle^{eq}, \bfp^{eq}(\bfx) $
	\end{algorithmic}\label{scft_algorithm}
\end{algorithm}

Algorithm \ref{scft_algorithm} shows the iterative procedure in the proposed self-consistent field theory formulation to obtain the equilibrium properties of a polarizable polymer chain. 

To initialize, i.e., guess the electric field for the initial step, we simply use $\phi (\bfx) = -\bfE_0 \cdot \bfx$, and use this to compute $\bfp$ from \eqref{eq:polarization_linear_form}.
To continue the numerical iteration from step $n$ to step $n+1$, we use $\bfp^n(\bfx)$, the polarization at iteration step $n$, to obtain the electric potential at the next iteration step, $\phi^{n+1}(\bfx)$, using:
\begin{equation}
    \nabla^2 \phi^{n+1}(\bfx)=  \frac{1}{\epsilon_0} \divergence \bfp^n(\bfx), \quad \text{ given } \phi^{n+1}(\bfx)= -\bfE_0\cdot \bfx \text{ on } \partial\Omega
\end{equation}
$\phi^{n+1}(\bfx)$ is used to obtain $w^{n+1}(\bfx, \bfu)$ using \eqref{eq:E_elec_potential_relation} and \eqref{eq:w_afo_E}. 
Using $w^{n+1}(\bfx, \bfu)$, we again solve \eqref{eq:q_PDE} and \eqref{eq:q_star_PDE} to compute $Q[w^{n+1}]$ and $\langle \hat{\rho}(\bfx, \bfu)  \rangle^{n+1} $ using \eqref{eq:Q} and \eqref{eq:rho_avg}, respectively.  
This in turn lets us compute $\bfp^{n+1}(\bfx)$ using \eqref{eq:avg_polarization_x_u_appendix} and \eqref{eq:avg_polarization_x}, which is used in the equation above to continue the iteration.
The self-consistent iteration procedure is continued until the energy term $-k_B T \log Q$ has converged, which we check by using the change in $Q$ across successive iterations.

\subsection{Finite Element Formulation}\label{sec:Finite Element Weak Formulation}

We use FEM to solve the PDEs \eqref{eq:q_PDE} and \eqref{eq:q_star_PDE} for the partial partition functions as in \cite{ackerman2017finite}.
We use FEniCS, an open-source FEM framework, for the numerical implementation \cite{AlnaesEtal2015,LoggEtal2012}. 
We work in 2 spatial dimensions (i.e., $\bfx = (x_1, x_2) \in \Omega \subset \mathbb{R}^2$) and restrict the unit orientation vector to the unit circle (i.e., it can be represented as $\bfu=(\cos \phi, \sin \phi)$, where $\phi \in [0, 2 \pi)$).
The configuration space in $(\bfx,\bfu)$ is 3-dimensional, enabling us to use standard FEM meshing and shape functions.

In terms of $q(x_1, x_2, \phi,s)$, we can rewrite \eqref{eq:q_PDE} as:
\begin{equation}\label{eq:q_PDE_weak_0}
	\frac{\partial q}{\partial s}= -w q - L_c \left( \cos{\phi} \frac{\partial q}{\partial x_1} + \sin{\phi} \frac{\partial q}{\partial x_2} \right) + \frac{L_c}{2 \lambda} \left( \frac{\partial^2 q}{\partial \phi^2} \right),
\end{equation}

The contour coordinate $s$ is treated as a time-like variable.
Derivatives with respect to $s$ in \eqref{eq:q_PDE} and \eqref{eq:q_star_PDE} are approximated using a Crank-Nicolson finite difference method.
We discretize in $s$ using a uniform discretization with $100$ steps along the chain contour.
We can then write:
\begin{equation}\label{eq:q_PDE_weak_1}
	\frac{q^{i+1}-q^i}{\Delta s}= \frac{f^{i+1}+f^i}{2},
	\quad \text{ with } f^i
	=
	-w q^i - L_c \left( \cos{\phi} \frac{\partial q^i}{\partial x_1} + \sin{\phi} \frac{\partial q^i}{\partial x_2} \right) + \frac{L_c}{2 \lambda} \left( \frac{\partial^2 q^i}{\partial \phi^2} \right)
\end{equation}
where the superscripts $i$ and $i+1$ represent the discretized quantities along $s$.

The domain in configuration space is discretized using first-order Lagrange family finite elements.
We use a mesh with $20 \times 40$ finite elements to discretize in $\bfx$ and $30$ finite elements to discretize in $\phi$, which is sufficiently refined that the quantities of interest are independent of the mesh.
The spatial mesh is finer around the chain ends,  and the Dirac delta functions in \eqref{eq:q_ini_condition} and \eqref{eq:q_star_ini_condition} are approximated as peaked Gaussians.
The mesh is uniform in the $\bfu$ discretization.

Following the usual FEM procedure, we first multiply \eqref{eq:q_PDE_weak_1} by a test function $v(x_1, x_2, \phi)$; second, integrate over $\bfx$ and $\bfu$; third, use integration-by-parts and the divergence theorem to convert the second derivatives $\frac{\partial^2 q}{\partial \phi^2}$ to a product of first derivatives; and, fourth, eliminate the boundary terms using the assumed Neumann boundary condition in $(x_1, x_2,\phi)$ to get the FEM weak form:
\begin{equation}\label{eq:WLC_weak_form}
	\begin{split}
	\int\limits_{\Tilde{\bfx},\phi} \left(  q^{i+1}v  + \frac{\Delta s}{2} w q^{i+1}v 
		+ \frac{ \Delta s}{2} \cos \phi \frac{\partial q^{i+1}}{\partial \Tilde{x}_1} v
		+ \frac{ \Delta s}{2} \sin \phi \frac{\partial q^{i+1}}{\partial \Tilde{x}_2} v
		+ \frac{ L_c \Delta s}{4 \lambda} \frac{\partial q^{i+1}}{\partial \phi} \frac{\partial v}{\partial \phi}  \right) \\
		=  
		\int\limits_{\Tilde{\bfx},\phi} \left(  q^{i} v - \frac{\Delta s}{2} w q^{i}v 
		- \frac{ \Delta s}{2} \cos \phi \frac{\partial q^{i}}{\partial \Tilde{x}_1} v
		- \frac{ \Delta s}{2} \sin \phi \frac{\partial q^{i}}{\partial \Tilde{x}_2} v
		- \frac{L_c \Delta s}{4 \lambda} \frac{\partial q^{i}}{\partial \phi} \frac{\partial v}{\partial \phi}  \right), 
	\end{split}
\end{equation} 
where $\Tilde{\bfx}=(\Tilde{x}_1, \Tilde{x}_2)= \left(\frac{x_1}{L_c}, \frac{x_2}{L_c} \right)$ is the nondimensional spatial coordinate.


\section{Results and Discussion}
\label{sec:results}

In this section, we apply the model to examine the effect of dipole-dipole interactions; specifically, we compare the electric field and dipole distributions with and without accounting for dipole interactions.
Then, we examine the effect of the orientation of the applied electric field, relative to the chain orientation, on the elastic response, the electric field, and the dipole distribution.

Various quantities are either nondimensionalized or rescaled, and will then be denoted by an overhead tilde $\tilde{\cdot}$. 
The chain is assumed to have $N=100$ polymer segments, each having length of $a=L_c/N$.
The length scales in the problem are nondimensionalized by chain contour length $L_c$.
The computational domain is chosen to be $-0.1 \leq \tilde{x}_1 \leq 0.1, -0.2 \leq \tilde{x}_2 \leq 0.2 $.
The electric field and polarization are both rescaled by dividing by $\left( \sqrt{ \frac{2 k_B T}{L_c^3}}  \right)$, i.e., $\tilde{\bfE} = \frac{\bfE}{\sqrt{ \frac{2 k_B T}{L_c^3}}}$ and $\tilde{\bfp}=\frac{\bfp}{\sqrt{ \frac{2 k_B T}{L_c^3}}}$.
For the polarizability tensor, we use $\beta_{\parallel}=1$ and $\beta_{\perp}=0.5$ following \cite{cohen2016electroelasticity}.
The angle between the applied electric field $\bfE_0$ with $\hat{\bfe}_1$ is denoted by $\theta_{\bfE_0}$. 

\subsection{Field and Dipole Distributions: Comparing with and without Dipole-dipole Interaction}

\begin{figure*}[ht!]
    \subfloat[Electric field distribution $\tilde{\bfE}$]{\includegraphics[width=0.32\textwidth]{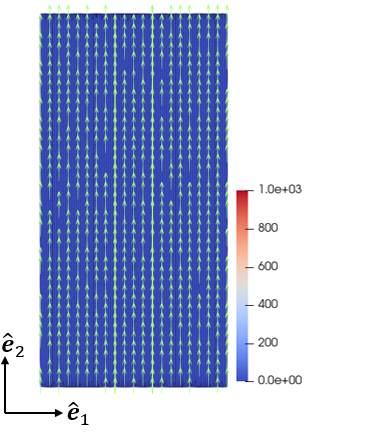}}
    \hfill
    \subfloat[Chain segment density $\tilde{\rho}$]{\includegraphics[width=0.32\textwidth]{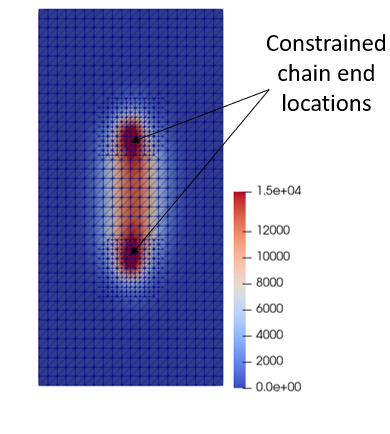}}
    \hfill
    \subfloat[Polarization distribution $\tilde{\bfp}$]{\includegraphics[width=0.32\textwidth]{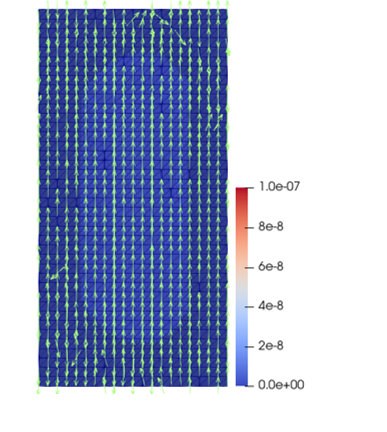}}
    \caption{
   	The baseline case with dipole-dipole interactions neglected, with externally applied electric field $\tilde{\bfE}_0=\hat{\bfe}_2$.
   	The arrows show the direction of the vector field, and the background color shows the magnitude.
   	(a) The electric field distribution $\tilde{\bfE}$, which is constant in space and equal to $\tilde{\bfE}_0$, as expected. 
   	(b) The chain segment density $\tilde{\rho}$. 
   	(c) The polarization distribution $\tilde{\bfp}$.
   	Both the electric field and polarization distributions are close to zero on the scale of the plot, which is chosen to allow comparison with the results when dipole-dipole interactions are included. 
     }
    \label{fig:no_dipole_dipole_interaction}
\end{figure*}

\begin{figure*}[ht!]
    \subfloat[Polarization distribution $\tilde{\bfp}$ for $\theta_{\bfE_0}=0$]{\includegraphics[width=0.47\textwidth]{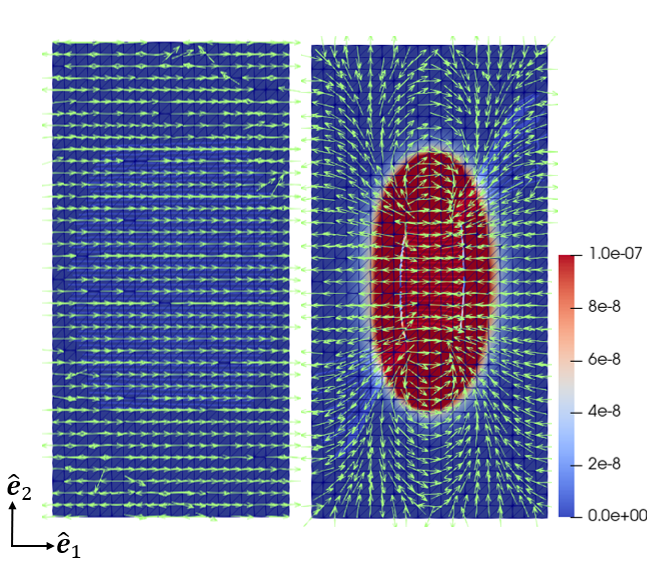}}
    \hfill
    \subfloat[Polarization distribution $\tilde{\bfp}$ for $\theta_{\bfE_0}=\pi/2$]{\includegraphics[width=0.47\textwidth]{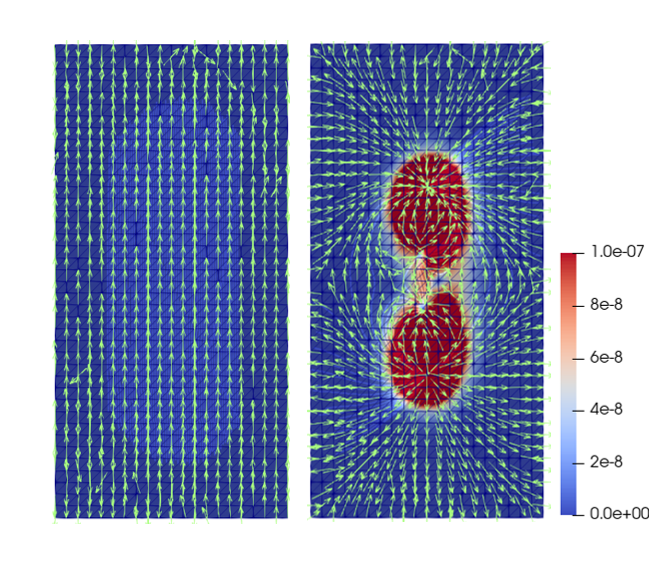}}
    \caption{
    Comparison of the polarization distribution $\tilde{\bfp}$, comparing without and with accounting for dipole-dipole interactions for different orientations of the applied electric field (a) $\theta_{\bfE_0}=0$, and (b) $\theta_{\bfE_0}= \pi/2$, both with $|\tilde{\bfE}_0|=1$. 
    In both (a) and (b), the left panel is without accounting for the interactions and the right panel is with accounting for the interactions. 	
    The polarization is at least one order of magnitude larger when dipole-dipole interactions are accounted for.
    The arrows show the direction of the vector field, and the background color shows the magnitude.}
\label{fig:comparison_of_no_dipole_interaction_and_full_dipole_interaction}
\end{figure*}

We fix the chain end-to-end vector length to $1.5 aN^{1/2}$ by fixing the chain ends at  $\pm 0.075 \ \bfe_2$ as shown in Figure \ref{fig:no_dipole_dipole_interaction}(b).
We apply a far-field electric field $\tilde{\bfE}_0=\hat{\bfe}_2$ by using the appropriate electric potential on the boundaries of the computational domain.

Figure \ref{fig:no_dipole_dipole_interaction} shows the electric field, chain segment density, and polarization for the baseline case when the dipole-dipole interactions are neglected.
We observe that the average segment density of the chain is highest at the constrained ends, and largely concentrated along the chain end-to-end vector. 
We also observe that the induced dipoles are essentially all oriented along $\tilde{\bfE}_0$.

Figure \ref{fig:comparison_of_no_dipole_interaction_and_full_dipole_interaction} shows a direct comparison of the polarization, $\tilde{\bfp}$, obtained without and with the dipole-dipole interactions for the extreme cases of the relative orientation between the applied field and the chain orientation.
We observe that the polarization with the interactions considered is at least an order of magnitude higher than the case that neglects the interactions.
Further, when the interactions are not considered, essentially all of the dipoles are oriented along $\tilde{\bfE}_0$, whereas there are much more complex dipole distributions when the interactions are considered.
When $\tilde{\bfE}_0$ is aligned with the chain end-to-end vector, the polarization is largely concentrated around the chain ends.
However, when $\tilde{\bfE}_0$ is orthogonal to the chain end-to-end vector, the polarization is relatively uniformly  distributed along the chain.

We conclude that the dipole-dipole interactions not only increase the polarization by orders of magnitude, but also changes the distribution, and depends strongly on the relative orientation of the applied electric field.

\subsection{Effect of the Orientation of the Electric Field Relative to the Chain}

\begin{figure*}[ht!]
    \subfloat[Electric potential $\tilde{\phi}$ for $\theta_{\bfE_0}=0$]{\includegraphics[width=0.33\textwidth]{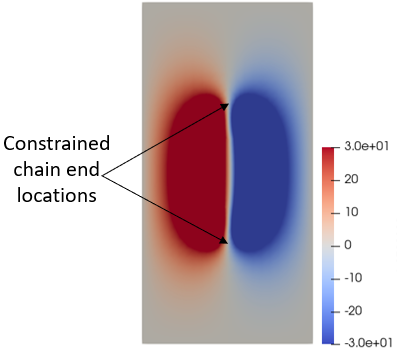}}
    \hfill
    \subfloat[Electric field $\tilde{\bfE}$ for $\theta_{\bfE_0}=0$]{\includegraphics[width=0.33\textwidth]{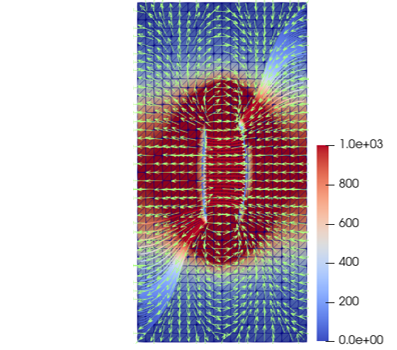}}
    \hfill
    \subfloat[Polarization $\tilde{\bfp}$ for $\theta_{\bfE_0}=0$]{\includegraphics[width=0.33\textwidth]{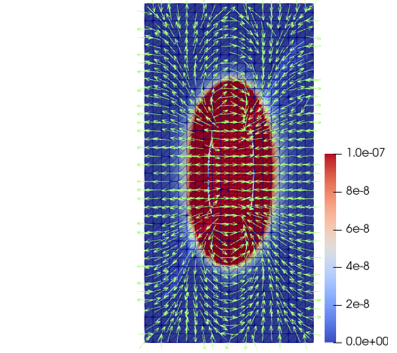}}
    \\
    \subfloat[Electric potential $\tilde{\phi}$ for $\theta_{\bfE_0}=\pi/2$]{\includegraphics[width=0.33\textwidth]{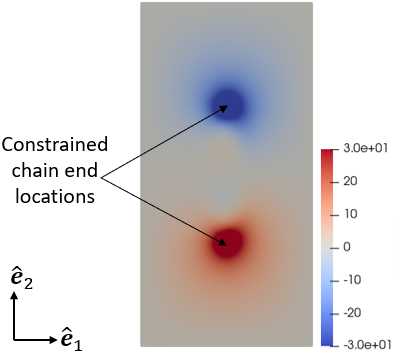}}
    \hfill
    \subfloat[Electric field $\tilde{\bfE}$ for $\theta_{\bfE_0}=\pi/2$]{\includegraphics[width=0.33\textwidth]{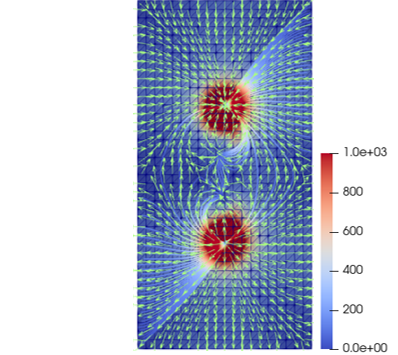}}
    \hfill
    \subfloat[Polarization $\tilde{\bfp}$ for $\theta_{\bfE_0}=\pi/2$]{\includegraphics[width=0.33\textwidth]{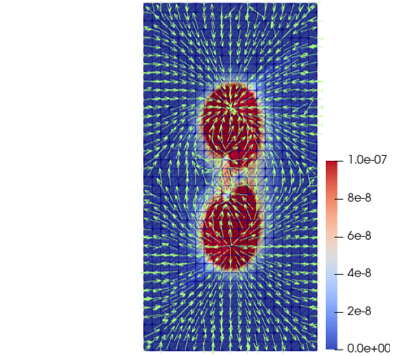}}
    \caption{
        Effect of orientation $\theta_{\bfE_0}$ of the applied electric field $\tilde{\bfE}_0$, with full accounting of the dipole-dipole interactions. 
   	We use $|\tilde{\bfE}_0(\bfx)|=1$. 
   	(a), (d) plot the electric potential;
   	(b), (e) plot the electric field;
   	and (c), (f)  plot the polarization, for $\theta_{\bfE_0}=0$ and $\theta_{\bfE_0}=\pi/2$ respectively.
   	The arrows show the direction of the vector field, and the background color shows the magnitude.
    }
\label{fig:effect_of_theta_Ea}
\end{figure*}

Next, we study the effect of the orientation of the applied electric field on the properties of the polymer chain. 
As in the earlier section, we fix the chain ends and fix the magnitude of the applied electric field $|\tilde{\bfE}_0|=1$, and study the effect of the field being orthogonal to the chain orientation compared to being aligned parallel to the chain orientation.

Figure \ref{fig:effect_of_theta_Ea} shows the electric potential, the electric field, and the polarization  for $\theta_{\bfE_0}=0$ and $\theta_{\bfE_0}=0\pi/2$.
When the applied electric field is orthogonal to the chain end-to-end vector ($\theta_{\bfE_0}=0$), we observe that the dipole distribution is approximately uniformly distributed along the chain.  
However, when the applied electric field is aligned with the chain end-to-end vector ($\theta_{\bfE_0}=\pi/2$), the dipole distribution is primarily concentrated near the chain ends.
The electric field distribution follows from the electric potential.
Further, we observe that when the applied electric field is orthogonal to the chain end-to-end vector, the electric field is roughly uniformly distributed along the chain.
However, when the applied electric field is aligned with the chain end-to-end vector, the electric field is largely concentrated near the chain ends.
We also observe that the local strength of the electric potential, the electric field, and the polarization is higher when the applied electric field is orthogonal to the chain end-to-end vector as compared to when it is aligned.

\subsection{Elastic Response}

\begin{figure}[ht!]
    \centering
    \includegraphics[width=0.98\textwidth]{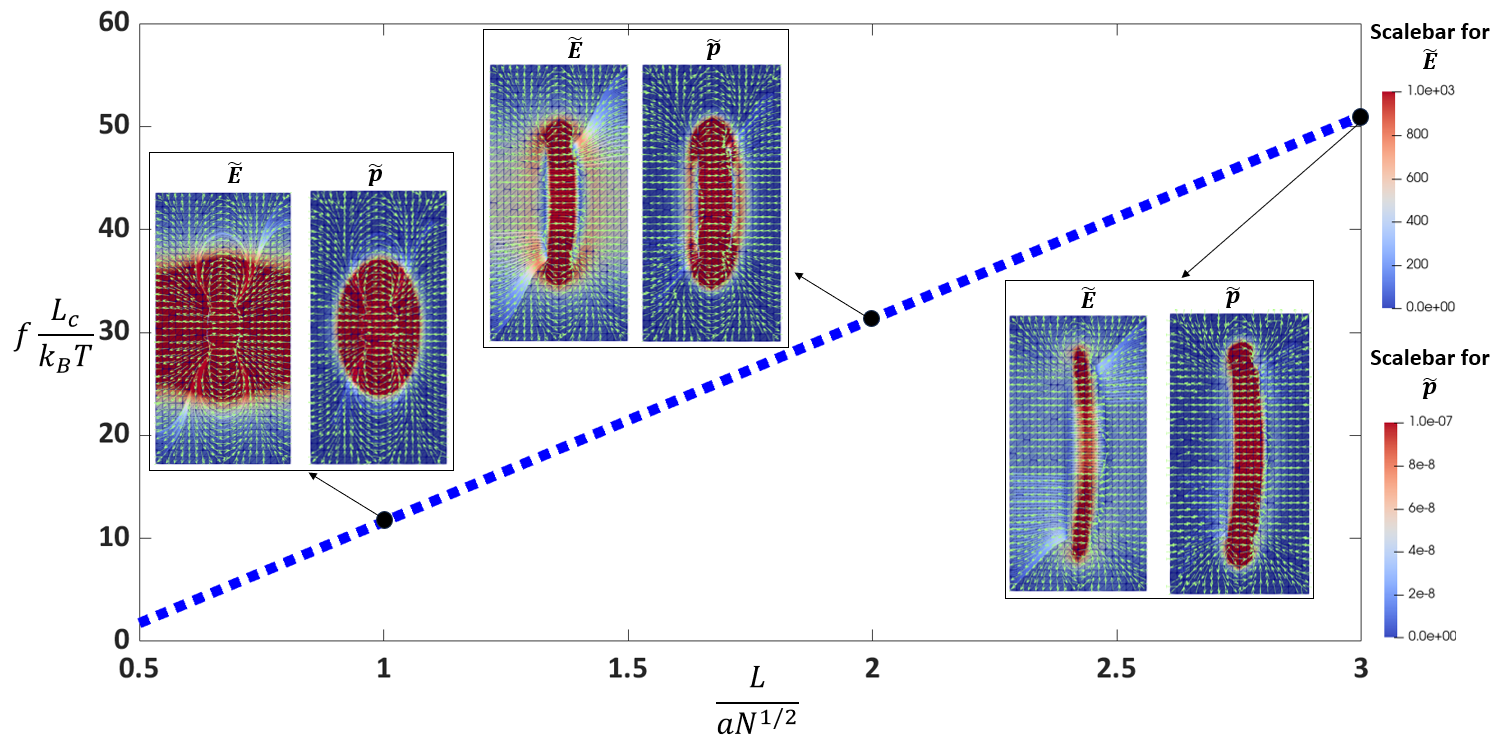}
    \caption{
        Elastic response (force against stretch) of the polymer chain for $\theta_{\bfE_0}=0$. 
        The insets show the electric field $\tilde{\bfE}$ and polarization $\tilde{\bfp}$ for $L=aN^{1/2}, 2aN^{1/2}, 3aN^{1/2}$.  
        The arrows show the direction of the vector field, and the background color shows the magnitude.
    }
\label{fig:force_vs_stretch_theta_0}
\end{figure}

\begin{figure}[ht!]
    \centering
    \includegraphics[width=0.98\textwidth]{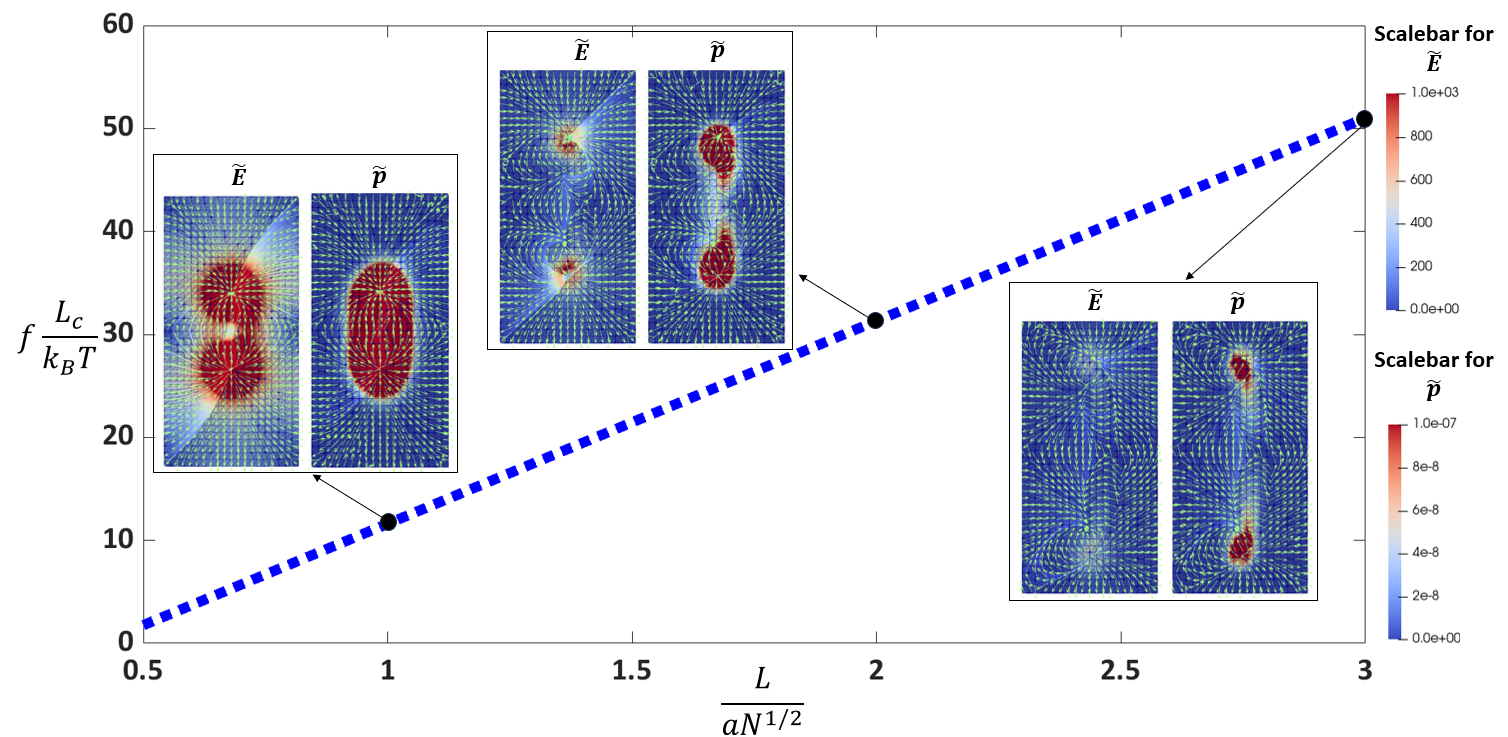}
    \caption{
        Elastic response (force against stretch) of the polymer chain for $\theta_{\bfE_0}=\pi/2$. 
        The insets show the electric field $\tilde{\bfE}$ and polarization $\tilde{\bfp}$ for $L=aN^{1/2}, 2aN^{1/2}, 3aN^{1/2}$.    
        The arrows show the direction of the vector field, and the background color shows the magnitude.
    }
\label{fig:force_vs_stretch_theta_90}
\end{figure}

To study the effect of chain stretch, we vary the length of the chain end-to-end vector, denoted $L$.
The total free energy of the polymer chain at equilibrium, $F$, is obtained as \cite{spencer2022self}:
\begin{equation}\label{eq:free_energy}
    F= -k_B T \log Q[w] -\frac{k_B T}{2} \int \dm\bfx \dm \bfu \ w(\bfx, \bfu) \langle \hat{\rho}(\bfx, \bfu) \rangle,
\end{equation}
The first term is the free energy of the non-interacting polymer chain in the external field $w(\bfx, \bfu)$, which double counts the electrical energy due to dipole-dipole interactions. 
The second term in \eqref{eq:free_energy} corrects for this double counting. 
We define the elastic force in the polymer chains by $f:=\frac{\partial F}{\partial L}$. 

Figures \ref{fig:force_vs_stretch_theta_0} and \ref{fig:force_vs_stretch_theta_90} show the elastic force as a function of $L$ for the applied electric field orthogonal and parallel to the chain orientation respectively. 
The insets in the figures show the electric field $\tilde{\bfE}$ and polarization $\tilde{\bfp}$. 
We observe that the elastic response is linear, which is consistent with the Gaussian nature of the polymer chain, even with the applied electric field at different orientations.
Further, the elastic force response is essentially unchanged even when we change the orientation of the applied electric field. 
However, the distributions of the electric field and polarization change very significantly with the direction of the applied electric field. 
When the applied electric field is orthogonal to the chain end-to-end vector ($\theta_{\bfE_0}=0$, Fig. \ref{fig:force_vs_stretch_theta_0}), we observe that as the chain is stretched, the distributions of the electric field and polarization remain roughly uniformly distributed along the chain, but are increasingly concentrated along the chain end-to-end vector with increasing stretch.  
However, when the applied electric field is aligned with the chain end-to-end vector ($\theta_{\bfE_0}=\pi/2$, Fig. \ref{fig:force_vs_stretch_theta_90}), the electric field and the polarization are primarily concentrated near the chain ends, with the concentration changing with increasing stretch.

\section{Concluding Remarks}
\label{sec:conclusions}

We have developed a statistical field theory framework for polarizable polymer chains that overcomes key limitations of existing approaches in accounting for the nonlocal dipole-dipole interactions between polymer segments. 
Our approach is applicable to general nonlinear polarization-electric field responses, by reformulating the nonlocal dipole-dipole interactions through the local PDE of electrostatics.
Regardless of the nonlinearity of the polarization-field response, the PDE constraint is linear and directly amenable to efficient numerical methods, such as boundary element methods that can account for unbounded domains \cite{dayal2007real} or very efficient Fourier methods \cite{peng2020effective}.
Our approach also enables the use of a more realistic ensemble that corresponds to a far-field applied electric field.

There are several possibilities for further development.
An immediate direction of study is to incorporate general nonlinear polarization-electric field responses that go beyond the linear anisotropic model studied here.
Another interesting direction would be to apply the framework developed here to study cross-linked polymer networks, e.g. following \cite{grasinger2023polymer}. 
This can provide key physical insights into the effects of inter-chain dipole-dipole interactions on the response of electromechanical polymer networks such as dielectric elastomers under external fields.
Yet another interesting possibility is to study polymers in confined settings with polarizable ambient media, which have been shown in other contexts to lead to unusual effects, e.g. \cite{robin2021modeling,dos2023modulation}.
Further, by including excluded volume interactions, it is possible to study the interplay between repulsive excluded volume interactions and attractive dipole-dipole interactions to tailor the functional properties of polymeric soft matter.  
Furthermore, an important aspect that we have not accounted for in our formulation is the role of fluctuations; recent developments in the statistical field theory provide tools to make progress in this direction \cite{ghosh2022semiflexible}.
Finally, the dipole response can be generalized to account for screening, rapid variations of the electric field, and so on \cite{martin2016statistical}.
In principle, these can be accounted for in our model --- screening by using the screened Poisson equation, for instance, and using gradients of the electric field in the monomer dipole response to account for quadrupole and higher moments --- but it is important to demonstrate these in practice.

\section*{}

\paragraph*{Software and Data Availability.}

The code developed for this work and the associated data are available at \\ 
\url{https://github.com/pkhandag/polarizable-polymer.git}

\paragraph*{Acknowledgments.}

We thank AFRL for hosting visits by Kaushik Dayal;
NSF (DMS 2108784, DMREF 1921857), BSF (2018183), and AFOSR (MURI FA9550-18-1-0095) for financial support;
and NSF for XSEDE computing resources provided by Pittsburgh Supercomputing Center.
This article draws from the doctoral dissertation of Pratik Khandagale at Carnegie Mellon University \cite{khandagale2022statistical}.


\end{document}

%% file: preamble.tex
\usepackage{isomath}
\usepackage{amsmath,amsthm}
\usepackage{amsbsy}
\usepackage{amssymb}
\usepackage{amscd}
\usepackage{amsfonts}
\usepackage{stmaryrd}
\usepackage{siunitx}
\usepackage{euscript}
\usepackage[utf8]{inputenc}
\usepackage[T1]{fontenc}
\usepackage{newtxtext} 
\everymath{\displaystyle}
\usepackage{exscale}
\usepackage{microtype}
\usepackage{hyperref}
\usepackage{booktabs}
\usepackage{algorithm}
\usepackage{algpseudocode}

\usepackage{graphicx}
\usepackage{boxedminipage}
\usepackage{calc}
\usepackage[usenames,dvipsnames]{xcolor}
\graphicspath{ {media/} }
\usepackage[caption=false,justification=centerlast]{subfig}

\usepackage{setspace}
\usepackage{enumitem}
\setitemize{noitemsep,topsep=0pt,parsep=0pt,partopsep=0pt}
\setenumerate{noitemsep,topsep=0pt,parsep=0pt,partopsep=0pt}
\setdescription{noitemsep,topsep=0pt,parsep=0pt,partopsep=0pt}

\usepackage{orcidlink}
\usepackage{siunitx}
\usepackage[small]{titlesec}

\titlespacing*{\section}{0pt}{12pt plus 4pt minus 2pt}{2pt plus 2pt minus 2pt}
\titlespacing*{\subsection}{0pt}{12pt plus 4pt minus 2pt}{2pt plus 2pt minus 2pt}
\titlespacing*\subsubsection{0pt}{12pt plus 4pt minus 2pt}{2pt plus 2pt minus 2pt}
\titlespacing*\paragraph{0pt}{12pt plus 4pt minus 2pt}{2pt plus 2pt minus 2pt}

\makeatletter

    \renewcommand*{\p@subsection}{}
    
    \renewcommand*{\p@subsubsection}{}
\makeatother

\input{definitions}

%% file: definitions.tex
\usepackage{isomath}
\usepackage{amsmath}
\usepackage{amssymb}
\usepackage{amscd}
\usepackage{amsfonts}
\usepackage{upgreek}

\theoremstyle{definition}

\AtEndEnvironment{remark}{\null\hfill\qedsymbol}

\AtEndEnvironment{example}{\null\hfill\qedsymbol}

\newcommand{\bfbeta}{\mathbold {\beta}}

\DeclareMathOperator{\divergence}{div}

\newcommand{\dm}{\ \mathrm{d}}

\newcommand{\bfe}{{\mathbold e}}

\newcommand{\bfp}{{\mathbold p}}

\newcommand{\bfr}{{\mathbold r}}

\newcommand{\bfu}{{\mathbold u}}

\newcommand{\bfx}{{\mathbold x}}

\newcommand{\bfE}{{\mathbold E}}

\newcommand{\bfI}{{\mathbold I}}

%% file: manuscript.bbl
\begin{thebibliography}{10}

\bibitem{brochu2012dielectric}
Paul Brochu and Qibing Pei.
\newblock Dielectric elastomers for actuators and artificial muscles.
\newblock {\em Electroactivity in polymeric materials}, pages 1--56, 2012.

\bibitem{kim2007electroactive}
Kwang~J Kim and Satoshi Tadokoro.
\newblock Electroactive polymers for robotic applications.
\newblock {\em Artificial Muscles and Sensors}, 23:291, 2007.

\bibitem{rosset2013self}
Samuel Rosset, Benjamin~M O’Brien, Todd Gisby, Daniel Xu, Herbert~R Shea, and
  Iain~A Anderson.
\newblock Self-sensing dielectric elastomer actuators in closed-loop operation.
\newblock {\em Smart Materials and Structures}, 22(10):104018, 2013.

\bibitem{leroy2020multimode}
Edouard Leroy, Ronan Hinchet, and Herbert Shea.
\newblock Multimode hydraulically amplified electrostatic actuators for
  wearable haptics.
\newblock {\em Advanced Materials}, 32(36):2002564, 2020.

\bibitem{ji2019autonomous}
Xiaobin Ji, Xinchang Liu, Vito Cacucciolo, Matthias Imboden, Yoan Civet, Alae
  El~Haitami, Sophie Cantin, Yves Perriard, and Herbert Shea.
\newblock An autonomous untethered fast soft robotic insect driven by
  low-voltage dielectric elastomer actuators.
\newblock {\em Science Robotics}, 4(37):eaaz6451, 2019.

\bibitem{cacucciolo2019stretchable}
Vito Cacucciolo, Jun Shintake, Yu~Kuwajima, Shingo Maeda, Dario Floreano, and
  Herbert Shea.
\newblock Stretchable pumps for soft machines.
\newblock {\em Nature}, 572(7770):516--519, 2019.

\bibitem{ware2016localized}
Taylor~H Ware, John~S Biggins, Andreas~F Shick, Mark Warner, and Timothy~J
  White.
\newblock Localized soft elasticity in liquid crystal elastomers.
\newblock {\em Nature communications}, 7(1):1--7, 2016.

\bibitem{babaei2021torque}
Mahnoush Babaei, Junfeng Gao, Arul Clement, Kaushik Dayal, and M~Ravi Shankar.
\newblock Torque-dense photomechanical actuation.
\newblock {\em Soft Matter}, 17(5):1258--1266, 2021.

\bibitem{huang2012giant}
Jiangshui Huang, Tiefeng Li, Choon Chiang~Foo, Jian Zhu, David~R Clarke, and
  Zhigang Suo.
\newblock Giant, voltage-actuated deformation of a dielectric elastomer under
  dead load.
\newblock {\em Applied Physics Letters}, 100(4), 2012.

\bibitem{erol2019microstructure}
Anil Erol, Saad Ahmed, Zoubeida Ounaies, and Paris von Lockette.
\newblock A microstructure-based approach to modeling electrostriction that
  accounts for variability in spatial locations of domains.
\newblock {\em Journal of the Mechanics and Physics of Solids}, 124:35--62,
  2019.

\bibitem{logrande2023dimensionally}
Kevin LoGrande, M~Ravi Shankar, and Kaushik Dayal.
\newblock A dimensionally-reduced nonlinear elasticity model for liquid crystal
  elastomer strips with transverse curvature.
\newblock {\em Soft Matter}, 19(45):8764--8778, 2023.

\bibitem{rogers2010materials}
John~A Rogers, Takao Someya, and Yonggang Huang.
\newblock Materials and mechanics for stretchable electronics.
\newblock {\em science}, 327(5973):1603--1607, 2010.

\bibitem{kim2009large}
Keun~Soo Kim, Yue Zhao, Houk Jang, Sang~Yoon Lee, Jong~Min Kim, Kwang~S Kim,
  Jong-Hyun Ahn, Philip Kim, Jae-Young Choi, and Byung~Hee Hong.
\newblock Large-scale pattern growth of graphene films for stretchable
  transparent electrodes.
\newblock {\em nature}, 457(7230):706--710, 2009.

\bibitem{khang2006stretchable}
Dahl-Young Khang, Hanqing Jiang, Young Huang, and John~A Rogers.
\newblock A stretchable form of single-crystal silicon for high-performance
  electronics on rubber substrates.
\newblock {\em Science}, 311(5758):208--212, 2006.

\bibitem{xu2013stretchable}
Sheng Xu, Yihui Zhang, Jiung Cho, Juhwan Lee, Xian Huang, Lin Jia, Jonathan~A
  Fan, Yewang Su, Jessica Su, Huigang Zhang, et~al.
\newblock Stretchable batteries with self-similar serpentine interconnects and
  integrated wireless recharging systems.
\newblock {\em Nature communications}, 4(1):1--8, 2013.

\bibitem{mckay2014dielectric}
Thomas~G McKay, Samuel Rosset, Iain~A Anderson, and Herbert Shea.
\newblock Dielectric elastomer generators that stack up.
\newblock {\em Smart Materials and Structures}, 24(1):015014, 2014.

\bibitem{han2019three}
Mengdi Han, Heling Wang, Yiyuan Yang, Cunman Liang, Wubin Bai, Zheng Yan, Haibo
  Li, Yeguang Xue, Xinlong Wang, Banu Akar, et~al.
\newblock Three-dimensional piezoelectric polymer microsystems for vibrational
  energy harvesting, robotic interfaces and biomedical implants.
\newblock {\em Nature Electronics}, 2(1):26--35, 2019.

\bibitem{erturk2011piezoelectric}
Alper Erturk and Daniel~J Inman.
\newblock {\em Piezoelectric energy harvesting}.
\newblock John Wiley \& Sons, 2011.

\bibitem{nan2018compliant}
Kewang Nan, Stephen~Dongmin Kang, Kan Li, Ki~Jun Yu, Feng Zhu, Juntong Wang,
  Alison~C Dunn, Chaoqun Zhou, Zhaoqian Xie, Matthias~T Agne, et~al.
\newblock Compliant and stretchable thermoelectric coils for energy harvesting
  in miniature flexible devices.
\newblock {\em Science advances}, 4(11):eaau5849, 2018.

\bibitem{kwak2020wireless}
Jean~Won Kwak, Mengdi Han, Zhaoqian Xie, Ha~Uk Chung, Jong~Yoon Lee, Raudel
  Avila, Jessica Yohay, Xuexian Chen, Cunman Liang, Manish Patel, et~al.
\newblock Wireless sensors for continuous, multimodal measurements at the skin
  interface with lower limb prostheses.
\newblock {\em Science translational medicine}, 12(574):eabc4327, 2020.

\bibitem{zhang2020water}
Qian Zhang, Qijie Liang, and John~A Rogers.
\newblock Water-soluble energy harvester as a promising power solution for
  temporary electronic implants.
\newblock {\em APL Materials}, 8(12):120701, 2020.

\bibitem{xu2014soft}
Sheng Xu, Yihui Zhang, Lin Jia, Kyle~E Mathewson, Kyung-In Jang, Jeonghyun Kim,
  Haoran Fu, Xian Huang, Pranav Chava, Renhan Wang, et~al.
\newblock Soft microfluidic assemblies of sensors, circuits, and radios for the
  skin.
\newblock {\em Science}, 344(6179):70--74, 2014.

\bibitem{kang2016bioresorbable}
Seung-Kyun Kang, Rory~KJ Murphy, Suk-Won Hwang, Seung~Min Lee, Daniel~V
  Harburg, Neil~A Krueger, Jiho Shin, Paul Gamble, Huanyu Cheng, Sooyoun Yu,
  et~al.
\newblock Bioresorbable silicon electronic sensors for the brain.
\newblock {\em Nature}, 530(7588):71--76, 2016.

\bibitem{kim2011epidermal}
Dae-Hyeong Kim, Nanshu Lu, Rui Ma, Yun-Soung Kim, Rak-Hwan Kim, Shuodao Wang,
  Jian Wu, Sang~Min Won, Hu~Tao, Ahmad Islam, et~al.
\newblock Epidermal electronics.
\newblock {\em science}, 333(6044):838--843, 2011.

\bibitem{jiao2014understanding}
Yang Jiao and Pinar Akcora.
\newblock Understanding the role of grafted polystyrene chain conformation in
  assembly of magnetic nanoparticles.
\newblock {\em Physical Review E}, 90(4):042601, 2014.

\bibitem{muthukumar2023physics}
Murugappan Muthukumar.
\newblock {\em Physics of Charged Macromolecules}.
\newblock Cambridge University Press, 2023.

\bibitem{nepal2023polymer}
Dhriti Nepal, Jeffrey Haines, and Richard~A Vaia.
\newblock Polymer nanocomposites: 35 years on.
\newblock {\em MRS Bulletin}, pages 1--11, 2023.

\bibitem{zolfaghari2020network}
Navid Zolfaghari, Pratik Khandagale, Michael~J Ford, Kaushik Dayal, and Carmel
  Majidi.
\newblock Network topologies dictate electromechanical coupling in liquid
  metal--elastomer composites.
\newblock {\em Soft Matter}, 16(38):8818--8825, 2020.

\bibitem{amjadi2016stretchable}
Morteza Amjadi, Ki-Uk Kyung, Inkyu Park, and Metin Sitti.
\newblock Stretchable, skin-mountable, and wearable strain sensors and their
  potential applications: a review.
\newblock {\em Advanced Functional Materials}, 26(11):1678--1698, 2016.

\bibitem{deng2014electrets}
Qian Deng, Liping Liu, and Pradeep Sharma.
\newblock Electrets in soft materials: Nonlinearity, size effects, and giant
  electromechanical coupling.
\newblock {\em Physical Review E}, 90(1):012603, 2014.

\bibitem{chen2021interplay}
Lingling Chen, Xu~Yang, Binglei Wang, Shengyou Yang, Kaushik Dayal, and Pradeep
  Sharma.
\newblock The interplay between symmetry-breaking and symmetry-preserving
  bifurcations in soft dielectric films and the emergence of giant
  electro-actuation.
\newblock {\em Extreme Mechanics Letters}, 43:101151, 2021.

\bibitem{zhao2021modeling}
Yongyi Zhao, Pratik Khandagale, and Carmel Majidi.
\newblock Modeling electromechanical coupling of liquid metal embedded
  elastomers while accounting stochasticity in 3d percolation.
\newblock {\em Extreme Mechanics Letters}, 48:101443, 2021.

\bibitem{liu2013giant}
Liping Liu and Pradeep Sharma.
\newblock Giant and universal magnetoelectric coupling in soft materials and
  concomitant ramifications for materials science and biology.
\newblock {\em Physical Review E}, 88(4):040601, 2013.

\bibitem{galipeau2013finite}
Evan Galipeau and Pedro~Ponte Casta{\~n}eda.
\newblock A finite-strain constitutive model for magnetorheological elastomers:
  magnetic torques and fiber rotations.
\newblock {\em Journal of the Mechanics and Physics of Solids},
  61(4):1065--1090, 2013.

\bibitem{zhang2017nonlinear}
Junshi Zhang, Hualing Chen, and Dichen Li.
\newblock Nonlinear dynamical model of a soft viscoelastic dielectric
  elastomer.
\newblock {\em Physical Review Applied}, 8(6):064016, 2017.

\bibitem{bar2004electroactive}
Yoseph Bar-Cohen.
\newblock {\em Electroactive polymer (EAP) actuators as artificial muscles:
  reality, potential, and challenges}, volume 136.
\newblock SPIE press, 2004.

\bibitem{grasinger2021flexoelectricity}
Matthew Grasinger, Kosar Mozaffari, and Pradeep Sharma.
\newblock Flexoelectricity in soft elastomers and the molecular mechanisms
  underpinning the design and emergence of giant flexoelectricity.
\newblock {\em Proceedings of the National Academy of Sciences}, 118(21), 2021.

\bibitem{kulkarni2023fluctuations}
Yashashree Kulkarni.
\newblock Fluctuations of active membranes with nonlinear curvature elasticity.
\newblock {\em Journal of the Mechanics and Physics of Solids}, 173:105240,
  2023.

\bibitem{hassan2023entropic}
Rubayet Hassan, Maria~Alejandra Garzon, Wei Gao, and Fatemeh Ahmadpoor.
\newblock Entropic pressure on fluctuating solid membranes.
\newblock {\em Journal of the Mechanics and Physics of Solids}, page 105523,
  2023.

\bibitem{ahmadpoor2013apparent}
F~Ahmadpoor, Q~Deng, LP~Liu, and P~Sharma.
\newblock Apparent flexoelectricity in lipid bilayer membranes due to external
  charge and dipolar distributions.
\newblock {\em Physical Review E}, 88(5):050701, 2013.

\bibitem{ahmadpoor2015flexoelectricity}
Fatemeh Ahmadpoor and Pradeep Sharma.
\newblock Flexoelectricity in two-dimensional crystalline and biological
  membranes.
\newblock {\em Nanoscale}, 7(40):16555--16570, 2015.

\bibitem{zelisko2017determining}
Matthew Zelisko, Fatemeh Ahmadpoor, Huajian Gao, and Pradeep Sharma.
\newblock Determining the gaussian modulus and edge properties of 2d materials:
  From graphene to lipid bilayers.
\newblock {\em Physical Review Letters}, 119(6):068002, 2017.

\bibitem{toupin1956elastic}
Richard~A Toupin.
\newblock The elastic dielectric.
\newblock {\em Journal of Rational Mechanics and Analysis}, 5(6):849--915,
  1956.

\bibitem{dorfmann2014nonlinear}
Luis Dorfmann and Ray~W Ogden.
\newblock {\em Nonlinear theory of electroelastic and magnetoelastic
  interactions}, volume~1.
\newblock Springer, 2014.

\bibitem{liu2018emergent}
Liping Liu and Pradeep Sharma.
\newblock Emergent electromechanical coupling of electrets and some exact
  relations—the effective properties of soft materials with embedded external
  charges and dipoles.
\newblock {\em Journal of the Mechanics and Physics of Solids}, 112:1--24,
  2018.

\bibitem{krichen2019liquid}
Sana Krichen, Liping Liu, and Pradeep Sharma.
\newblock Liquid inclusions in soft materials: Capillary effect, mechanical
  stiffening and enhanced electromechanical response.
\newblock {\em Journal of the Mechanics and Physics of Solids}, 127:332--357,
  2019.

\bibitem{li2015geometrically}
Xiaobao Li, Liping Liu, and Pradeep Sharma.
\newblock Geometrically nonlinear deformation and the emergent behavior of
  polarons in soft matter.
\newblock {\em Soft matter}, 11(41):8042--8047, 2015.

\bibitem{darbaniyan2019designing}
Faezeh Darbaniyan, Kaushik Dayal, Liping Liu, and Pradeep Sharma.
\newblock Designing soft pyroelectric and electrocaloric materials using
  electrets.
\newblock {\em Soft matter}, 15(2):262--277, 2019.

\bibitem{fox2008dynamic}
JW~Fox and NC~Goulbourne.
\newblock On the dynamic electromechanical loading of dielectric elastomer
  membranes.
\newblock {\em Journal of the Mechanics and Physics of Solids},
  56(8):2669--2686, 2008.

\bibitem{li2023constitutive}
Yali Li and Nakhiah~C Goulbourne.
\newblock Constitutive formulations for intrinsic anisotropy in soft
  electroelastic materials.
\newblock {\em Scientific Reports}, 13(1):14712, 2023.

\bibitem{furer2022homogenization}
Joshua Furer and Pedro~Ponte Casta{\~n}eda.
\newblock Homogenization, macroscopic instabilities and domain formation in
  magnetoactive composites: Theory and applications.
\newblock {\em Journal of the Mechanics and Physics of Solids}, 169:105081,
  2022.

\bibitem{cohen2016electroelasticity}
Noy Cohen, Kaushik Dayal, and Gal deBotton.
\newblock Electroelasticity of polymer networks.
\newblock {\em Journal of the Mechanics and Physics of Solids}, 92:105--126,
  2016.

\bibitem{cohen2016electromechanical}
Noy Cohen and Gal deBotton.
\newblock Electromechanical interplay in deformable dielectric elastomer
  networks.
\newblock {\em Physical review letters}, 116(20):208303, 2016.

\bibitem{grasinger2020statistical}
Matthew Grasinger and Kaushik Dayal.
\newblock Statistical mechanical analysis of the electromechanical coupling in
  an electrically-responsive polymer chain.
\newblock {\em Soft Matter}, 16(27):6265--6284, 2020.

\bibitem{grasinger2021architected}
Matthew Grasinger and Kaushik Dayal.
\newblock Architected elastomer networks for optimal electromechanical
  response.
\newblock {\em Journal of the Mechanics and Physics of Solids}, 146:104171,
  2021.

\bibitem{grasinger2021nonlinear}
Matthew Grasinger, Carmel Majidi, and Kaushik Dayal.
\newblock Nonlinear statistical mechanics drives intrinsic electrostriction and
  volumetric torque in polymer networks.
\newblock {\em Physical Review E}, 103(4):042504, 2021.

\bibitem{grasinger2022statistical}
Matthew Grasinger, Kaushik Dayal, Gal deBotton, and Prashant~K Purohit.
\newblock Statistical mechanics of a dielectric polymer chain in the force
  ensemble.
\newblock {\em Journal of the Mechanics and Physics of Solids}, 158:104658,
  2022.

\bibitem{itskov2019electroelasticity}
Mikhail Itskov, Vu~Ngoc Khi{\^e}m, and Sugeng Waluyo.
\newblock Electroelasticity of dielectric elastomers based on molecular chain
  statistics.
\newblock {\em Mathematics and Mechanics of Solids}, 24(3):862--873, 2019.

\bibitem{treloar1975physics}
Leslie Ronald~George Treloar.
\newblock {\em The physics of rubber elasticity}.
\newblock Oxford University Press, USA, 1975.

\bibitem{kuhn1946statistical}
Werner Kuhn and Franz Gr{\"u}n.
\newblock Statistical behavior of the single chain molecule and its relation to
  the statistical behavior of assemblies consisting of many chain molecules.
\newblock {\em Journal of polymer science}, 1(3):183--199, 1946.

\bibitem{weiner2012statistical}
Jerome~Harris Weiner.
\newblock {\em Statistical mechanics of elasticity}.
\newblock Courier Corporation, 2012.

\bibitem{marckmann2006comparison}
Gilles Marckmann and Erwan Verron.
\newblock Comparison of hyperelastic models for rubber-like materials.
\newblock {\em Rubber chemistry and technology}, 79(5):835--858, 2006.

\bibitem{khandagale2023statistical}
Pratik Khandagale, Timothy Breitzman, Carmel Majidi, and Kaushik Dayal.
\newblock Statistical field theory for nonlinear elasticity of polymer networks
  with excluded volume interactions.
\newblock {\em Physical Review E}, 107(6):064501, 2023.

\bibitem{fredrickson2007computational}
Glenn~H Fredrickson.
\newblock Computational field theory of polymers: opportunities and challenges.
\newblock {\em Soft Matter}, 3(11):1329--1334, 2007.

\bibitem{fredrickson2002field}
Glenn~H Fredrickson, Venkat Ganesan, and Fran{\c{c}}ois Drolet.
\newblock Field-theoretic computer simulation methods for polymers and complex
  fluids.
\newblock {\em Macromolecules}, 35(1):16--39, 2002.

\bibitem{fredrickson2006equilibrium}
Glenn Fredrickson.
\newblock {\em The equilibrium theory of inhomogeneous polymers}, volume 134.
\newblock Oxford University Press on Demand, 2006.

\bibitem{martin2016statistical}
Jonathan~M Martin, Wei Li, Kris~T Delaney, and Glenn~H Fredrickson.
\newblock Statistical field theory description of inhomogeneous polarizable
  soft matter.
\newblock {\em The Journal of chemical physics}, 145(15):154104, 2016.

\bibitem{shen2017electrostatic}
Kevin Shen and Zhen-Gang Wang.
\newblock Electrostatic correlations and the polyelectrolyte self energy.
\newblock {\em The Journal of chemical physics}, 146(8), 2017.

\bibitem{jiang2018density}
Jian Jiang, Valeriy~V Ginzburg, and Zhen-Gang Wang.
\newblock Density functional theory for charged fluids.
\newblock {\em Soft Matter}, 14(28):5878--5887, 2018.

\bibitem{zhuang2018statistical}
Bilin Zhuang and Zhen-Gang Wang.
\newblock Statistical field theory for polar fluids.
\newblock {\em The Journal of Chemical Physics}, 149(12), 2018.

\bibitem{marshall2014atomistic}
Jason Marshall and Kaushik Dayal.
\newblock Atomistic-to-continuum multiscale modeling with long-range
  electrostatic interactions in ionic solids.
\newblock {\em Journal of the Mechanics and Physics of Solids}, 62:137--162,
  2014.

\bibitem{ackerman2017finite}
David~M Ackerman, Kris Delaney, Glenn~H Fredrickson, and Baskar
  Ganapathysubramanian.
\newblock A finite element approach to self-consistent field theory
  calculations of multiblock polymers.
\newblock {\em Journal of Computational Physics}, 331:280--296, 2017.

\bibitem{AlnaesEtal2015}
M.~S. Alnaes, J.~Blechta, J.~Hake, A.~Johansson, B.~Kehlet, A.~Logg,
  C.~Richardson, J.~Ring, M.~E. Rognes, and G.~N. Wells.
\newblock The {FEniCS} project version 1.5.
\newblock {\em Archive of Numerical Software}, 3, 2015.

\bibitem{LoggEtal2012}
A.~Logg, {K.-A.} Mardal, G.~N. Wells, et~al.
\newblock {\em Automated Solution of Differential Equations by the Finite
  Element Method}.
\newblock Springer, 2012.

\bibitem{spencer2022self}
Russell~KW Spencer, Bae-Yeun Ha, and Nima Saeidi.
\newblock Self-consistent field theory of chiral nematic worm-like chains.
\newblock {\em The Journal of Chemical Physics}, 156(11), 2022.

\bibitem{dayal2007real}
Kaushik Dayal and Kaushik Bhattacharya.
\newblock A real-space non-local phase-field model of ferroelectric domain
  patterns in complex geometries.
\newblock {\em Acta materialia}, 55(6):1907--1917, 2007.

\bibitem{peng2020effective}
Xiaoyao Peng, Dhriti Nepal, and Kaushik Dayal.
\newblock Effective response of heterogeneous materials using the recursive
  projection method.
\newblock {\em Computer Methods in Applied Mechanics and Engineering},
  364:112946, 2020.

\bibitem{grasinger2023polymer}
Matthew Grasinger.
\newblock Polymer networks which locally rotate to accommodate stresses,
  torques, and deformation.
\newblock {\em Journal of the Mechanics and Physics of Solids}, 175:105289,
  2023.

\bibitem{robin2021modeling}
Paul Robin, Nikita Kavokine, and Lyd{\'e}ric Bocquet.
\newblock Modeling of emergent memory and voltage spiking in ionic transport
  through angstrom-scale slits.
\newblock {\em Science}, 373(6555):687--691, 2021.

\bibitem{dos2023modulation}
AP~Dos~Santos, F~Jim{\'e}nez-{\'A}ngeles, A~Ehlen, and M~Olvera De~La~Cruz.
\newblock Modulation of ionic conduction using polarizable surfaces.
\newblock {\em Physical Review Research}, 5(4):043174, 2023.

\bibitem{ghosh2022semiflexible}
Ashesh Ghosh, Quinn MacPherson, Zhen-Gang Wang, and Andrew~J Spakowitz.
\newblock Semiflexible polymer solutions. ii. fluctuations and frank elastic
  constants.
\newblock {\em The Journal of Chemical Physics}, 157(15), 2022.

\bibitem{khandagale2022statistical}
Pratik~Dinkar Khandagale.
\newblock {\em Statistical Field Theory for Polymer Networks and
  Electro-mechanical Polymers: Accounting for Nonlocal Interactions}.
\newblock PhD thesis, Carnegie Mellon University, 2022.

\end{thebibliography}
